# Gross-Pitaevskii Model for Nonzero Temperature Bose-Einstein Condensates

Kestutis Staliunas

Physikalisch Technische Bundesanstalt, 38116 Braunschweig, Germany

tel.: +49-531-5924482,  Fax: +49-531-5924423,  E-mail: Kestutis.Staliunas@PTB.DE

**Abstract**

Momentum distributions and temporal power spectra of nonzero temperature Bose-Einstein condensates are calculated using a Gross-Pitaevskii model. The distributions are obtained for micro-canonical ensembles (conservative Gross-Pitaevskii equation) and for grand-canonical ensembles (Gross-Pitaevskii equation with fluctuations and dissipation terms). Use is made of an equivalence between statistics of the solutions of conservative Gross Pitaevskii and dissipative complex Ginzburg-Landau equations. In all cases the occupation numbers of modes follow a $\langle N_k \rangle \propto k^{-2}$ dependence, which corresponds in the long wavelength limit ($k \to 0$) to Bose-Einstein distributions. The temporal power spectra are of $1/f^a$ form, where: $a = 2 - D/2$ with $D$ the dimension of space.

**Introduction**

Bose-Einstein condensates (BECs) are usually described by a Gross-Pitaevskii Equation (GPE) [1]. It is an open question presently in which way the nonzero temperature BECs are best described. We attempt here a GPE description: we describe conservative condensates as by the usual GPE, and also nonconservative condensates (e.g. atom lasers) by a GPE with fluctuation- dissipation terms, which corresponds to the contact with a thermal bath. One way of description is a systematic quantum mechanical derivation of GPE models. Such attempts [2,3], however, did not lead to any simple and clear model. We therefore start from a phenomenological model, and then justify it by its results (i.e. by the momentum distributions following from the phenomenological model).

Nonzero temperature BECs are usually decomposed into a condensed and a noncondensed part, following the traditional approach in superfluidity. The dynamics of the condensed part is governed by a GPE, coupled to the equations governing the dynamics of the noncondensed part [4]. This description, although generally accepted, and although yielding results compatible with experiments, is somewhat artificial. The separation of the atoms into condensed and uncondensed ones is ad hoc. More consistent would be a unified description either in terms of a wave- or a particle picture. This article uses a description in the wave picture as it is adequate with zero temperature condensates.

The motivation for such a description is nonlinear optics, where partially coherent fields in lasers are well described in a wave picture [5]. (The lasers are indeed nonzero temperature condensates of photons.) The source of incoherence (e.g. quantum noise) is simulated by additional stochastic Langevin forces. The success of the laser theory in describing laser dynamics by stochastic wave equations [5] suggests, that nonzero temperature condensates should be describable analogously by modified GPEs.

Attempts for description of nonzero temperature BECs in a wave picture have been made by deriving the modified GPEs quantum mechanically [2,3], as well as by introducing dissipative terms phenomenologically [6,7]. However presently the situation is still confusing: the systematic quantum mechanical derivations did not lead to any simple and clear model, and the simple phenomenological models are not proven to yield correct statistical distributions. Recent attempts [7] to calculate numerically the equilibrium distributions of fields described by a phenonenologically introduced dissipative GPE lead to controversial results: the mode intensities in the long wavelength limit ($k \to 0$) were found to follow a $\langle N_k \rangle \propto k^{-1}$ distribution, which is, however, not consistent with the well known theorem, that (not trapped) condensates are only stable for $D \geq 3$ dimensions of space [4]. This inconsistency is perhaps due to the fact that analytical and numerical treatment of the GPE, as of a conservative equation, is complicated, e.g. in that the distributions computed are converging only very slowly towards equilibrium distributions, as is shown below.

Here statistical distributions of GPE solutions both for the conservative and weakly dissipative cases are calculated. Instead of straightforward calculation, which numerically is a difficult task due to the conservative nature of the GPE, an alternative method is used. First, it is shown that the equilibrium distributions for the conservative GPE are identical to the distributions for the stochastic Complex Ginzburg-Landau Equation (CGLE) with real valued coefficients. The CGLE is a dissipative and variational counterpart of the GPE, and can be solved easier numerically and analytically. Then the distributions recently calculated for a CGLE with real valued coefficients [8] are also those of the GPE, due to this equivalence. Second, the analysis is extended to grand-canonical ensembles described by a GPE with fluctuation and dissipation terms corresponding to contact with a thermal bath.

The case of two spatial dimensions (2D) is investigated numerically in this letter. A common opinion is that the GPE in two (and more than two) spatial dimensions is nonintegrable. Therefore ergodicity of solutions of the conservative GPE can be expected (but is not guaranteed). The ergodicity is an essential assumption for the analytical results, since the equilibrium distributions calculated there have a meaning for ergodic systems only. The ergodicity allows to use the term of "temperature" for the excitations in the GPE. Therefore numerical calculations were performed in order to investigate the convergence of the distributions to the equilibrium. The calculations, as shown below, show convergence (although slow) towards equilibrium, evidencing (weak) ergodicity of the GPE in 2D.

**Conservative case (GPE)**

It is well known that the GPE, derived to describe the zero temperature condensates [8]:

$$\frac{\partial \psi}{\partial t} = i \cdot \left( \psi - |\psi|^2 \psi + \nabla^2 \psi \right) \qquad (1)$$

is a Hamiltonian one, which can be rewritten in the form:

$$\frac{\partial \psi}{\partial t} = i \cdot \frac{dH(\{\psi\})}{d\psi^*} \qquad (2)$$

with a Hamilton function:

$$H(\{y\}) = \int \left( -|y|^2 + \frac{|y|^4}{2} + |\nabla y|^2 \right) d\vec{r} \qquad (3)$$

where the figure brackets denote that the Hamilton function depends on the order parameter $y$ at every location of space. Of coordinate space $\vec{r}$, or of momentum (Fourier) space $\vec{k}$: $y(k) = \int y(\vec{r}) \cdot \exp(ik\vec{r}) \cdot d\vec{r}$.

We use the GPE (1) to describe excited, or nonzero temperature, ensembles, assuming ergodicity and equipartition for the solutions of the GPE in two or higher dimensions of space. Then the equilibrium distribution formally follows from the equipartition assumption:

$$P(\{y\}) = Z \cdot \exp\left( \frac{-H(\{y\})}{T} \right) \qquad (4)$$

Here $P(\{y\})$ is the joint probability distribution that the order parameter $y$ at every location of space (either coordinate or momentum space) is of a given amplitude and phase. $Z$ is the partition function, and T is the temperature of the equilibrium distribution, proportional to the total energy of the field (2).

The stationary distribution function $P(\{y\})$ allows in principle to calculate the spatial spectra (mode intensities):

$$<N_k> = \left\langle |y(\vec{k})|^2 \right\rangle = \int |y(\vec{k})|^2 P(\{y\}) d\{y\} \qquad (5)$$

where $P(\{y\})$ is the joint probability distribution in the momentum space, and the integration in (5) is performed over all of the configuration space (over the complex amplitudes of all spatial modes).

Equations (3-5) allow formally to calculate the momentum distribution at equipartition, however, practically they are of little use. The calculation of integrals is possible only under some approximations, as e.g. in the Bogoliubov theory [9].

**Dissipative case (CGLE)**

Our treatment is based on the observation, that the CGLE with real valued coefficients:

$$\frac{\partial y}{\partial t} = y - |y|^2 y + \nabla^2 y \qquad (6)$$

is a variational one, which can be rewritten in the form:

$$\frac{\partial y}{\partial t} = -\frac{dF(\{y\})}{dy^*} \qquad (7)$$

with variational potential (Lyapunov function) exactly coinciding with the Hamilton function of GPE (1): $F(\{y\}) \equiv H(\{y\})$. If one adds stochastic Langevin terms to (6):

$$\frac{\partial y}{\partial t} = y - |y|^2 y + \nabla^2 y + \Gamma(\vec{r}, t) \tag{8}$$

where $\Gamma(\vec{r},t)$ is an additive noise, $\delta$ - correlated in space and time, and of intensity $T$: $<\Gamma(\vec{r}_1,t_1)\cdot\Gamma(\vec{r}_2,t_2)> = 2T\cdot\delta(\vec{r}_1-\vec{r}_2)\delta(t_1-t_2)$, one can write the Fokker-Plank equation for the temporal evolution of the joint probability distribution $P(\{y\})$ of (8):

$$\frac{\partial P(\{y\})}{\partial t} = \frac{1}{2}\int\left(\frac{\partial}{\partial y}\left(\frac{dF}{dy^*}\right)+\frac{\partial}{\partial y^*}\left(\frac{dF}{dy}\right)+T\left(\frac{\partial}{\partial y^*}\frac{\partial}{\partial y}+\frac{\partial}{\partial y}\frac{\partial}{\partial y^*}\right)\right)P(\{y\})\cdot d\vec{r} \tag{9}$$

which at detailed balance has the stationary solution:

$$P(\{y\}) = N\cdot\exp\left(\frac{-F(\{y\})}{T}\right) \tag{10}$$

Comparing (10) with (4) one can conclude that the stationary distributions for the stochastic CGLE at detailed balance is equivalent to the equilibrium distributions for the GPE. In this way, the stochastic properties of solutions of the stochastic CGLE (e.g. the distributions in the momentum space), are identical to those of the GPE in equilibrium.

The spatial power spectra of the stochastic CGLE with real valued coefficients (8) were recently calculated [8] both analytically and numerically. They were shown to obey a $\langle N_k \rangle \propto k^{-2}$ power law, independent of the number of spatial dimensions of the system (the 1D, 2D and 3D cases were numerically analyzed). In this way the average mode intensities of the GPE at equilibrium are proven to follow the same $\langle N_k \rangle \propto k^{-2}$ dependence.

**Intermediate case (dissipative GPE)**

The analogy between purely dissipative and purely conservative systems holds also for intermediate cases. Indeed the stochastic complex CGLE:

$$\frac{\partial y}{\partial t} = (c+i)\cdot\left(y - |y|^2 y + \nabla^2 y\right) + c^{1/2}\cdot\Gamma(\vec{r},t) \tag{11}$$

whose deterministic part can be written in form:

$$\frac{\partial y}{\partial t} = (c+i)\cdot\frac{dF(\{y\})}{dy^*} \tag{12}$$

has also the Fokker-Planck equation:

$$\frac{\partial P(\{y\})}{\partial t} = \frac{c}{2}\int\left(\frac{\partial}{\partial y}\left(\frac{dF(\{y\})}{dy^*}\right)+\frac{\partial}{\partial y^*}\left(\frac{dF(\{y\})}{dy}\right)+T\left(\frac{\partial}{\partial y^*}\frac{\partial}{\partial y}+\frac{\partial}{\partial y}\frac{\partial}{\partial y^*}\right)\right)P(\{y\})\cdot d\vec{r}$$

$$-\frac{1}{2}\int\left(\frac{\partial}{\partial y}\left(\frac{dF(\{y\})}{dy^*}\right)-\frac{\partial}{\partial y^*}\left(\frac{dF(\{y\})}{dy}\right)\right)P(\{y\})\cdot d\vec{r} \tag{13}$$

whose stationary, detailed balance solution for arbitrary values of $c$ coincides with the above stationary solutions (4) and (10) for the purely conservative and purely dissipative cases respectively.

We note that the conservative case $c=0$ can be also formally described by Fokker-Plank equation (13). The detailed balance solution of (13) in this case corresponds to equipartition distribution.

It is essential that the dissipative part in (11) is proportional to the conservative part. Only in this case can the Fokker-Plank equation be (formally) solved. Mathematically the proportionality between conservative and dissipative part means that the equation has a Lyapunov potential. Physically this proportionality was introduced phenomenologically [6] using the argument that only such form of the dissipative part ensures the relaxation of condensates towards equilibrium.

**Distributions in momentum space**

We performed numerical integrations of (11) with values of $c$ ranging from zero (purely conservative case) to infinity (purely dissipative case), in order to confirm the above equivalence, and gain information about the assumptions such as ergodicity. Fig.1 shows a series of calculated mode intensities, averaged in time, which indicate the same $\langle N_k \rangle \propto k^{-2}$ dependence for all cases. Initial conditions for calculations in dissipative cases were spatially homogeneous distributions, which converged rapidly to the stationary ones. In the conservative case we started the calculations with a stochastic term and a nonzero dissipative part $c \neq 0$, waited until stationary distributions were reached, and then "switched off" the noise and dissipative terms and continued calculating with $c=0$.

The numerical integration was performed on a grid of (128*128), which restricts the magnitude of available wavenumbers to roughly two decades. In order to obtain spectra extending over a larger range of spatial wavenumbers a multiscale technique was applied, as described in [8]: the calculations were performed using different temporal and spatial steps, and the spatial spectra were combined into one plot. This technique allowed to obtain spatial spectra extending over four decades, as shown in Fig.1.

In all three cases shown in Fig.1.a the $\langle N_k \rangle \propto k^{-2}$ dependence is evident. However, the data points are much more scattered in the conservative case of $c=0$. Although the numerical data points lie around the $\langle N_k \rangle \propto k^{-2}$ line, the convergence towards the line (toward equilibrium state of GPE in case of two spatial dimensions) was slow. Below the convergence towards equilibrium is analyzed in detail (Fig.3).

Fig.1.b. shows the normalized spatial power spectrum $k^2 S(k)$ for the case $c=1$. The spectrum is flat at large spatial frequency $k \gg 1$, where amplitude and phase fluctuations contribute equally, and at small spatial frequencies $k \ll 1$, where amplitude fluctuations are negligibly small compared with phase fluctuations, in accordance with results for the purely dissipative case [8]. The same distributions are observed in the weakly dissipative case $c=0.1$, and also, although less clearly (due to strongly scattered data points), in the conservative case.

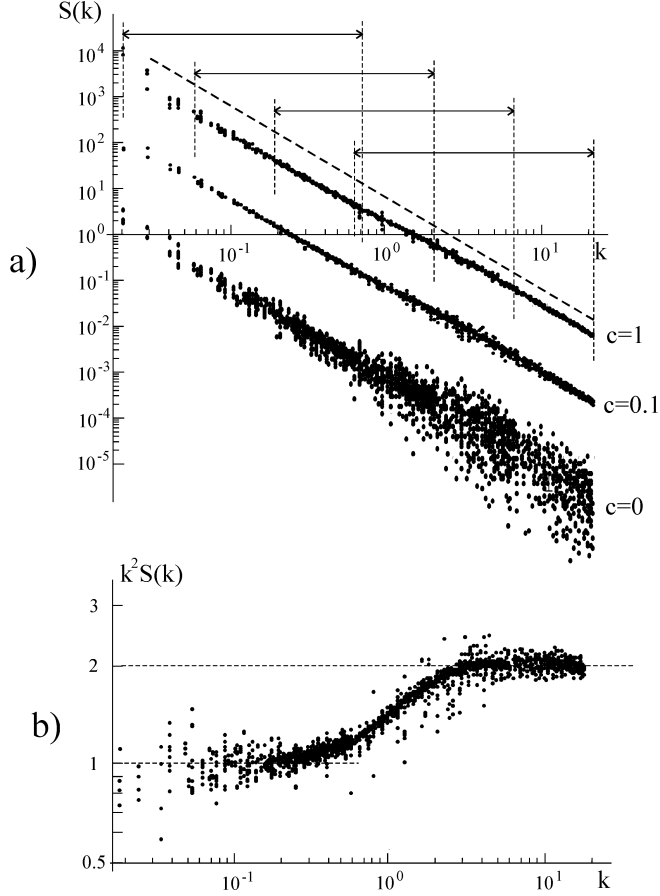

*Fig.1.*

*a) Spatial power spectra (momentum distributions) as obtained by numerical integration of stochastic CGLE (11) for different values of dissipation. The distributions are in log-log representation, and are shifted arbitrarily in vertical direction. The grid of (128*128) was used. The dashed line with slope $k^{-2}$ is to guide the eye. The arrows indicate the parts of the spectra obtained by separate calculations, which then were combined into one plot.*

*b) Normalized spatial power spectra $k^2 S(k)$ for the case $c = 1$.*

**Temporal power spectra**

The above analytical and numerical results prove the equivalence of the spatial power spectra for the dissipative and conservative cases. The equivalence between temporal power spectra does not follow from the equivalence of stationary probability distributions. However, numerical integration of (11) shows that the temporal power spectra are also equivalent for the dissipative and conservative cases. A series of power spectra of amplitude and phase fluctuations is shown in Fig.2, which indicate this equivalence. The temporal power spectra for the stochastic CGLE ($c = \infty$) as found in [8], are of $1/w^a$ form, where: $a = 2 - D/2$. The spectrum of amplitude fluctuations is Lorenz-like $1/(2p + w)^a$ saturating at small frequencies and dropping with the same exponent $a$ for large frequencies. The same dependence was found numerically for conservative and intermediate cases, as Fig.2. shows: the main characteristics (1/f-like $1/w^a$ form for phase fluctuations, and Lorenz-like $1/(2p + w)^a$ form for amplitude fluctuations) are clearly visible in all calculated cases. However, we could not prove analytically the identity between temporal power spectra for the two (conservative and dissipative) limits.

Like for the momentum distributions, the convergence of the temporal power spectra with the increasing averaging time was also slow. The data points remain strongly scattered, however, evidently lie around the Lorenz-like power spectra: $1/w$ for phase fluctuations, and $1/(2 + w)$ for amplitude fluctuations.

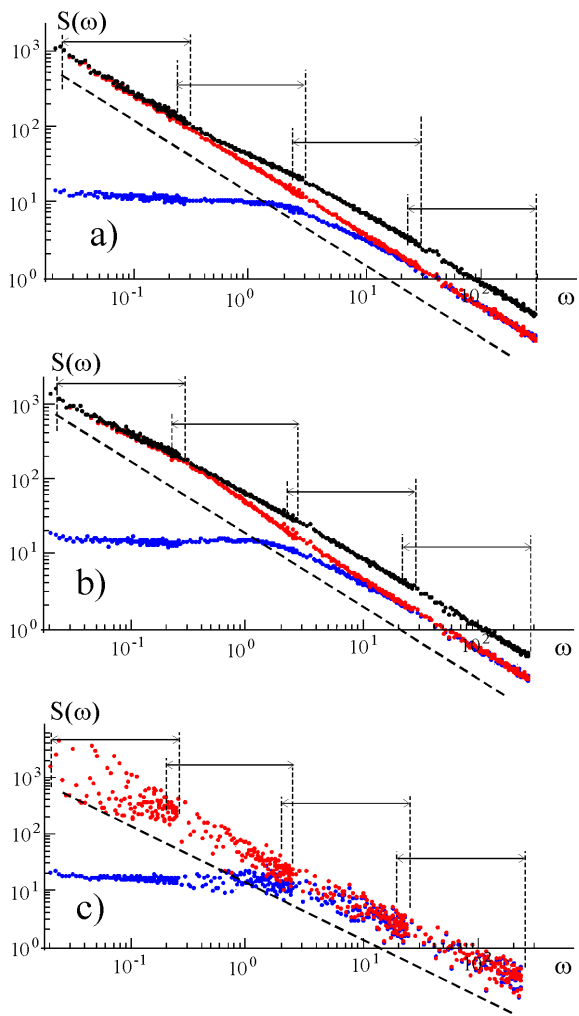

*Fig.2. Temporal power spectra as obtained by numerical integration of stochastic CGLE (11) for different values of dissipation: a) for $c = 1$, b) for $c = 0.1$, and c) for $c = 0$. The distributions are in log-log representation. Spectra of intensity fluctuations are shown by blue points (the lowest curves) and the spectra of phase fluctuations are shown by red points (the middle curves). Black points (the upper curves in a) and b) cases) show the total spectrum. The dashed lines with the $\omega^{-1}$ slope are to guide the eye.*

**Convergence**

The convergence of the temporal power spectra and of momentum distribution towards equilibrium was slow in conservative cases, as Fig.1, and Fig.2. indicate. The data points remain strongly scattered despite long integration (and averaging) times. In order to explore the character of convergence we performed a corresponding numerical study: we fitted the averaged momentum distributions, and calculated the variance as the mean square displacement of the data points from the fit curve. We followed the decrease of the variance with increasing time of averaging.

An ergodic stochastic process should lead to a linear decrease of the variance with increasing averaging time: $\sigma^2 \propto t_{averaging}^{-1}$. Such behaviour was observed for the weakly dissipative case of CGLE (11), as indicated by the lower curve (the line in log-log representation) in Fig.3. However, in the conservative case the variance is decreasing, with increasing averaging time, much slower than for ergodicity, as the upper curve in Fig.3 indicates. Moreover, the variance is not decreasing following a power law, but more likely following a logarithmic law. This indicates that the GPE in 2D is weakly ergodic. We do not have a clear explanation for weak ergodicity. One guess would be that the spatial distributions in 2D space are in general partially one-dimensional, and partially two dimensional. (E.g. if starting numerical integration with the initial distributions depending only one coordinate, the dynamics remains 1D). This mixture of 1D (integrable, nonergodic) and 2D (nonintegrable, ergodic) dynamics could cause the weak ergodicity observed numerically.

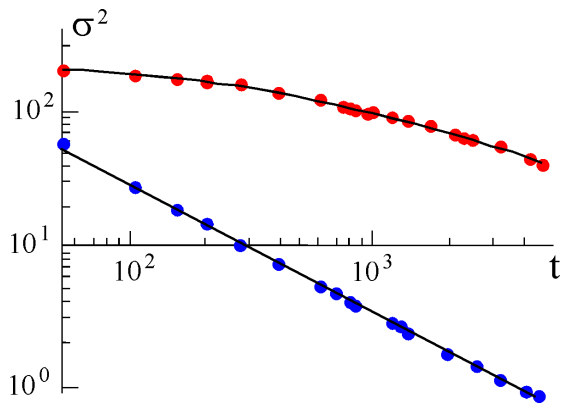

*Fig.3.* The variance of the averaged momentum distribution for weakly dissipative ($c = 0.1$, the lower line) and conservative ($c = 0$, the upper line) cases of stochastic CGLE (11) depending on integration (and averaging) time.

Another explanation for weak ergodicity follows from some indications that the 2D GPE may be only weakly nonintegrable [10]. As shown in [10] the vortices of the defocusing GPE (or equivalently, of the defocusing nonlinear Schrödinger equation) are solitons standing the Painleve test and the numerical test of elasticity of collisions. However, a large ensemble of such elastic solitons behaves obviously chaotically, as every billiards in 2D. These results lead to a conclusion of weak nonintegrability of 2D nonlinear Schrödinger equation in [10], which is compatible with weak ergodicity observed here.

**Conclusions**

The equivalence between stationary distributions in conservative systems and in corresponding dissipative systems is shown. In particular the conservative GPE describes micro-canonical ensembles of BECs, where the condensate is prepared at a fixed temperature. The nonconservative CGLE (11) with nonzero parameter $c$ describes grand-canonical ensembles of BECs (or atom lasers) where the temperature, and the density of particles in condensate is determined by the thermal bath.

The equivalence between distributions for conservative and dissipative cases, as proven in this paper, allows to transfer results between these two limiting cases. In particular, it results that the momentum distributions in micro-canonical and grand-canonical ensembles of GPE at equipartition are the same as in a purely dissipative CGLE, and obey the $\langle N_k \rangle \propto k^{-2}$ dependence. The dependence of $\langle N_k \rangle \propto k^{-2}$ coincides with the Bose Einstein distributions in the long wavelength limit. The Gaussian tail of Bose-Einstein distributions in the short wavelength limit was not obtained for GPE models.

The equivalence of temporal power spectra is shown only numerically: in all conservative, dissipative and intermediate cases is the 1/f character of the spectra shown, and a clear dominance of phase fluctuations over amplitude fluctuations in the long wavelength limit is predicted.

The results are in contradiction to those in [7], where a $\langle N_k \rangle \propto k^{-1}$ distribution in momentum space is shown. One possible explanation of this is the difference of the preparation of the condensates in [7] and in our case. We prepared the condensate by integrating the corresponding dissipative equations, and thus started with the $\langle N_k \rangle \propto k^{-2}$ distribution, which persists in the further evolution. In [7] the condensate was prepared by starting with the $\langle N_k \rangle \propto k^{-1}$ dependence, which also persisted during the evolution. Comparing both results one may conclude that the thermalisation is slow in GPE equation. This conclusion is supported by our numerical results in conservative cases, where only weak convergence towards the equilibrium was found with increasing integration time.

## Acknowledgement

The work has been supported by Sonderforschugsbereich 407 of Deutsche Forschungsgemeinschaft. Discussions with C.O.Weiss, M.Lewenstein are acknowledged.